# Investigation of lightweight acoustic curtains for mid-to-high frequency noise insulations

Sanjay Kumar, Jie Wei Aow, Wong Dexuan, and Heow Pueh Lee

Department of Mechanical Engineering, National University of Singapore, 9 Engineering Drive 1, Singapore 117575, Singapore

Corresponding author's email: mpesanj@nus.edu.sg (SK)

**Abstract**

The continuous surge of environmental noise levels has become a vital challenge for humanity. Earlier studies have reported that prolonged exposure to loud noise may cause auditory and non-auditory disorders. Therefore, there is a growing demand for suitable noise barriers. Herein, we have investigated several commercially available curtain fabrics' acoustic performance, potentially used for sound insulation purposes. Thorough experimental investigations have been performed on PVC coated polyester fabrics' acoustical performances and 100 % pure PVC sheets. The PVC-coated polyester fabric exhibited better sound insulation properties, particularly in the mid-to-high frequency range (600-1600 Hz) with a transmission loss of about 11 to 22 dB, while insertion loss of > 10 dB has been achieved. Also, the acoustic performance of multi-layer curtains has been investigated. These multi-layer curtains have shown superior acoustic properties to that of single-layer acoustic curtains.

**Introduction**

In the past few decades, environmental noise has increased substantially and posed a considerable risk to the residents and commuters. The community's unpleasant noise emissions are produced by outdoor activities such as machine operations, traffic, constructions, transportations, industries, and indoor activities such as household equipment, loud speech, music, etc. Together, these generated noises can create a highly unpleasant and unhealthy situation and exceed the prescribed noise level for a healthy environment. Noise pollution is severe in urban cities worldwide. It is caused mainly by traffic, along with densely traveled roads. Equivalent sound pressure levels for 24 hours generally range from 75–80 dBA. During peak hours, it often exceeds the average values. The World Health Organization (WHO) has considered a high noise level one of the most crucial factors that directly adversely impacts human beings [1]. Previous research findings support the WHO claims that the prolonged exposure to these unpleasant noises with exceeding limits may induce negative physiological and psychological impacts on human health. The prolonged exposure to noise in the community has severe implications for residents, such as hearing impairment, sleep disturbance, speech intelligibility, mental illness, social and behavioral effects like an annoyance, hypertension, agitation, psychological complications, like increased perceived work pressure, prone to committing work errors, stress, headache, increased fatigue levels, emotional exhaustion, and burnout [2]. In this perspective, effective action to limit and control exposure to environmental noise is needed. In line with the WHO, the National Environment Agency (NEA), Singapore has specified the maximum permissible noise level standards for different noise sources. The set values for construction work noise in residential premises, hospitals, schools, institutions of higher learning, homes for aged sick, etc., are 60 dBA ($L_{eq}$= 12 hrs) during the daytime and 50 dB(A) ($L_{eq}$ = 12 hrs) at night. Whereas, in the residential buildings located less than 150m from the construction site, the values are 75 dBA ($L_{eq}$= 12 hrs) during the daytime and 55 dB(A) ($L_{eq}$ = 1 hr) at night. Similarly permitted industrial noise limits for noise Sensitive Premises are 60 dBA and 50 dB(A), for residential Premises 65 dBA and 55 dB(A), and the commercial premises 70 dBA and 60 dB(A) in the daytime and night (11 PM-7 AM), respectively [3].

In recent years considerable efforts have been attempted for the community noise mitigations to ensure a healthy living environment. Some architectural design solutions include installing high-performance sound-absorbing panels, sonic crystals, noise barriers, etc.[4–9]. Acoustic curtains are widely used for soundproofing, mainly in residential households (at windows and doors), construction sites, and commercial buildings. These curtains are made of various materials such as polyesters, cotton, multi-layer fabrics, polyvinyl chloride (PVC), etc. These materials are lightweight, durable, foldable, affordable, and easily installable at desired locations. In recent times, the acoustic performance of these curtain materials has been widely explored. The research findings revealed the potentials of acoustic materials for noise mitigation applications. Wang et al.[10] measured the sound transmission loss of laminated mica-filled PVC composites. Tang et al.[11] studied the acoustic absorption properties of woven fabrics. Zhang et al.[12] investigated the sound insulation performance of a fiber-based flexible porous composite with a dual-gradient structure. The composite was made of nonwoven fabric, which was coated on one side. The presented dual-gradient system resulted in broadband sound insulation. Li et al.[13] investigated the acoustic absorption capabilities of woven structure fabrics. Yu and Kang[14] studied the acoustic properties such as transmission loss and reverberation time of the building acoustic materials (e.g., glass curtain). The environmental impact was also accessed for these acoustic materials. Segura-Alcaraz et al.[15] measured the sound absorption coefficient of multi-layer fabrics made of a combination of woven fabric and nonwoven fabrics. Tang et al.[16] measured the sound absorption properties of a multi-layered structure composed of nonwoven materials and polyethylene membranes.

Most of these published works were material-based investigations, and their acoustic properties were measured using a laboratory-based impedance tube system. However, the reports on the acoustic measurement of real-size curtains are minimal. Pieren et al.[17] assessed the sound absorption coefficients of the woven fabrics made of synthetic materials (polyester).

The experimentation was performed in the reverberation room, and textiles were suspended in two settings: flat curtains backed by the wall and free-hanging folded curtains. Therefore, a holistic approach is still needed to select suitable acoustic curtain types for effective noise mitigations.

Herein, we performed the acoustical analysis of several potential acoustic curtain materials made of polyvinyl chloride (PVC), polystyrene, polypropylene, and other fabrics. The acoustic performance of these materials has been comprehensively investigated on small-scale (impedance tube method) and large-scale levels (reverberation room method). The whole analysis was performed in three different strategies. First, all specimens' sound transmission loss performance was evaluated using the transfer matrix method-based four microphone impedance tube system. Based on their performances, the potential materials were sorted out, and then the reverberation room method was used to measure the insertion loss of these curtains with the size of 1m×1m. In the third strategy, acoustic characteristics of multi-layer curtains were explored using the reverberation room method.

## Materials and methods

### Materials

The materials for the fabrication of noise barriers were carefully selected based on several parameters like durability, weight, and water-repellent features. Nine different materials were procured and investigated; namely, TANGO curtain, elephant mat type I and type II (95% polypropylene and 5% Polyester, purchased from Nippon Home), woolen felt (soft and stiff), 100% pure PVC sheet (transparent), PVC coated polyester (PE) fabric (Feicheng Hengfeng Plastic. Co), PVC and polyester-based textured soft liner (GRIP by *Ladelle Pty Ltd*.), and antibacterial 100% polyester hospital curtain (Shaoxing Dairui®). All the test specimens were cut into square pieces and carefully weighed to calculate the surface density. The surface mass density was calculated by multiplying mass density and the sample thickness. The material specimen details have been listed in Table 1.

**Table 1.** Material Specifications Summary.

| Material types | Thickness (mm) | Surface Density (kg/m^2) |
|---|---|---|
| Woolen felt (woven, Soft) | 1.235 | 0.213 |
| Woolen felt (woven, Stiff) | 1.922 | 0.252 |
| TANGO curtain | 0.57 | 0.224 |
| 100% polyester hospital curtain | 0.6 | 0.229 |
| Elephant mat (Type I) | 2.276 | 0.318 |
| Elephant mat (Type II) | 1.682 | 0.366 |
| Textured soft liner (GRIP) | 1.65 | 0.644 |
| PVC coated polyester (PE) fabric | 0.89 | 1.135 |
| 100% pure PVC sheet | 1.012 | 1.216 |

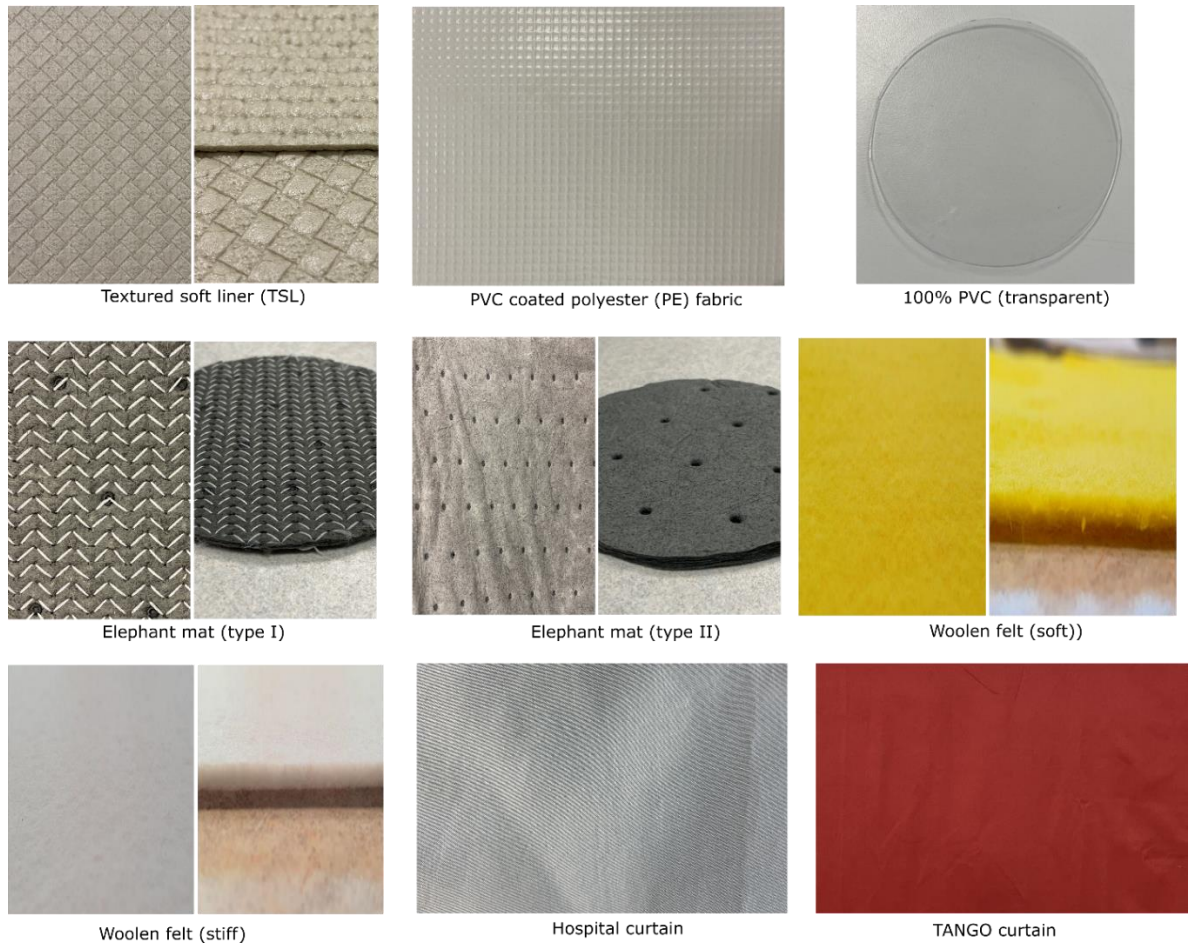

**Figure 1.** Photographs of the materials used for the experimental investigations.

**Acoustical measurements**

We have used two measurement techniques for the characterization of the acoustic performance of the curtains [18]. These methods are described as follows.

**Impedance tube method**

A standard sound impedance tube (BSWA®) was used for the transmission loss measurements. The specifications of the impedance tube were designed as per the British standard EN ISO 10534-2:2001. Figure 2(a) shows the schematic diagram of the four-microphone impedance tube system. For the transmission loss measurements, a four-microphone setup was adopted where two pairs of microphones (BSWA MPA416) were installed at the upstream and downstream of the test specimen to acquire sound waves pressure data. Two 3D printed polylactide ring-shaped sample holders (outer dia. 99.8 mm, inner dia. 79 mm, and thickness 14 mm) was used to mount thin and flexible curtains in the impedance tube. The test specimens of diameter 99.8 mm were attached to the sample holder using double-sided adhesive tape. To ensure uniform tension of the specimens, these were laid flat before affixing to the contacting surface of the holder, as suggested by Weilnau et al.[19].

Moreover, the position of the sample holder assembly was varied by rotating it along the circumference of the tube in an interval of 45 degrees, and STL was measured at each position. Further, five repetitions of STL measurements were taken at each position to minimize the possible experimental errors by sample placement. The average of these values was reported in the present study.

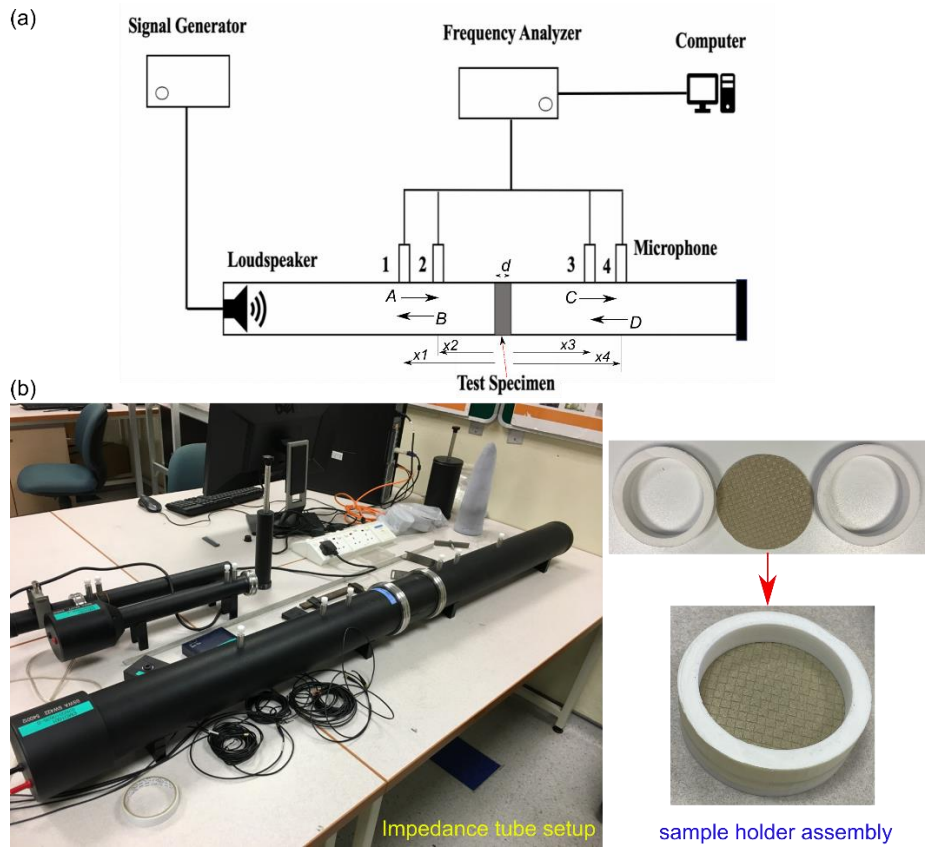

**Figure 2.** (a) Schematic diagram of the four-microphone impedance tube system. (b) Photographs of the impedance tube with sample holder arrangements.

The transfer matrix function was used to calculate the sound transmission loss of the specimen in the four-microphone impedance system. For a plane wave propagating sound source, the acoustic pressure upstream and downstream of the sample is expressed as

$$P_{up} = Ae^{-jkx} + Be^{jkx} \tag{1}$$

$$P_{down} = Ce^{-jkx} + De^{jkx} \tag{2}$$

The sound pressure at four microphone locations can be written as

$$P_1 = Ae^{-jkx_1} + Be^{jkx_1} \tag{3}$$

$$P_2 = Ae^{-jkx_2} + Be^{jkx_2} \tag{4}$$

$$P_3 = Ce^{-jkx_3} + De^{jkx_3} \tag{5}$$

$$P_4 = Ce^{-jkx_4} + De^{jkx_4} \tag{6}$$

On solving,

$$A = \frac{j(P_1 e^{jkx_2} - P_2 e^{jkx_1})}{2 \sin k \, (x_1 - x_2)} \tag{7}$$

$$B = \frac{j(P_3 e^{jkx_4} - P_4 e^{jkx_3})}{2 \sin k \, (x_3 - x_4)} \tag{8}$$

$$C = \frac{j(P_2 e^{-jkx_1} - P_1 e^{-jkx_2})}{2 \sin k \, (x_1 - x_2)} \tag{9}$$

$$D = \frac{j(P_4 e^{-jkx_3} - P_3 e^{-jkx_4})}{2 \sin k \, (x_3 - x_4)} \tag{10}$$

A 2×2 transfer matrix can model the pressure and velocity at the entry and exit point of the sample,

$$\begin{bmatrix} P \\ V \end{bmatrix}_{x=0} = \begin{bmatrix} T_{11} & T_{12} \\ T_{21} & T_{21} \end{bmatrix} \begin{bmatrix} P \\ V \end{bmatrix}_{x=d} \tag{11}$$

At x = 0,

$$P|_{x=0} = A + B; \quad V|_{x=0} = \frac{A - B}{\rho_0 c} \tag{12}$$

At x = d,

$$P|_{x=d} = Ce^{-jkd} + De^{jkd}; \quad V|_{x=d} = \frac{Ce^{-jkd} - De^{jkd}}{\rho_0 c} \tag{13}$$

To solve the matrix (11), two additional equations are required,

$$\begin{bmatrix} P_1 & P_2 \\ V_1 & V_2 \end{bmatrix}_{x=0} = \begin{bmatrix} T_{11} & T_{12} \\ T_{21} & T_{21} \end{bmatrix} \begin{bmatrix} P_1 & P_2 \\ V_1 & V_2 \end{bmatrix}_{x=d} \tag{14}$$

Here, $T_{11} = T_{22}$ (Reciprocity and symmetry) and $T_{11}T_{22} - T_{12}T_{21} = 1$

For a perfectly anechoic termination, D = 0, and considering incident plane wave with unit amplitude,

$$P|_{x=0} = 1 + R_a; \quad V|_{x=0} = \frac{1 - R_a}{\rho_0 c} \tag{15}$$

$$P|_{x=d} = T_a e^{-jkd}; \quad V|_{x=d} = \frac{T_a e^{-jkd}}{\rho_0 c} \tag{16}$$

where $R_a = B/A$ and $T_a = C/A$ are the reflection and transmission coefficients, respectively.

By combining Eqs. (14)-(16), the transfer matrix elements for a sample can be evaluated. Also, the sound reflection and transmission coefficients can be estimated from the following equations.

$$T_a = \frac{2e^{jkd}}{T_{11} + \frac{T_{12}}{\rho_0 c} + T_{21} + T_{22}} \tag{17}$$

$$R_a = \frac{T_{11} + \frac{T_{12}}{\rho_0 c} - \rho_0 c T_{21} - T_{22}}{T_{11} + \frac{T_{12}}{\rho_0 c} + T_{21} + T_{22}} \tag{18}$$

Also, normal surface impedance is given as

$$Z_a = \frac{T_{11} + \frac{T_{12}}{\rho_0 c}}{T_{21} + \frac{T_{22}}{\rho_0 c}} \tag{19}$$

The equations (17)-(19) are valid for anechoic termination.

For a rigid backing plate arrangement, $V|_{x=d} = 0$, and reflection coefficient ($R_h$) is given as

$$R_h = \frac{T_{11} - \rho_0 c T_{21}}{T_{11} + \rho_0 c T_{21}} \tag{20}$$

Finally, the transmission loss is calculated from the expression,

$$STL = 10 \log \frac{1}{|T_a|^2} \tag{21}$$

**Reverberation room method**

Large-scale specimens' acoustical performance was measured in the ASTM standard reverberation chamber at the National University of Singapore acoustics laboratory. The enclosure consisted of a source and receiver room having volumes of 65.5 m³ and 81.1 m³, respectively. The difference in room volumes was about 19.2%, fulfilling the volume difference requirement of at least 10%, as specified in ASTM E90 [20]. The surface areas of the source and the receiver rooms were 97.2 m² and 109.5 m², respectively. The rooms were separated by a partition wall with a central opening (0.995 X 0.995 X 0.104 m³) to mount specimens. The temperature and relative humidity on the day of testing was 23.7°C and 66.3%, respectively. Test specimens were tight-fitted carefully within the central opening area to ensure there would be no sound leakages. Two wooden frames were used to firmly hold the thin curtains, ensuring the flatness of these specimens. As the samples had a smaller thickness relative to the thickness of the partition wall opening, wooden frames were delicately designed as a filler for the samples to be held firmly by the pre-installed toggle clamps. Figure 3(a) shows the PVC coated curtain installation setup photographs for sound acoustical measurements. Gaskets were placed between the wooden frames' contacting surfaces and toggle clamps to minimize potential sound leakages.

*Insertion loss calculation*

A separate setup was arranged for the measurement of average sound pressure levels (SPLs) in the source room $L_s(f)$ and in the receiving room, $L_r(f)$ as per ASTM E90. Figure 3(b, c) shows the reverberation room's schematic representation for the SPL measurement. A continuous pink noise was played into the source room using the two omnidirectional loudspeakers (Yamaha DXR15), driven by separate random noise generators and amplifiers. The loudspeakers were placed at the trihedral corners of the room to excite room modes effectively. Two calibrated microphones in the source room and the receiver room were used to capture the sound levels in the respective rooms. All the measurements were performed in all one-third-octave bands with nominal mid-band frequencies specified in ANSI S1.11 from 100 to 5000 Hz. Six experiments were performed for each specimen, and their average values were considered in transmission loss calculation. The insertion loss (IL) was evaluated by the following relations $IL = L_{ro}(f) - L_{rs}(f)$. Where $L_{ro}(f)$ and $L_{rs}(f)$ represent the sound pressure level measured in the receiver room before and after the sample installation at the partition wall. A repeatability test was performed at each frequency to check with the required confidence interval for transmission loss measurements as suggested by standard ASTM E90 [20].

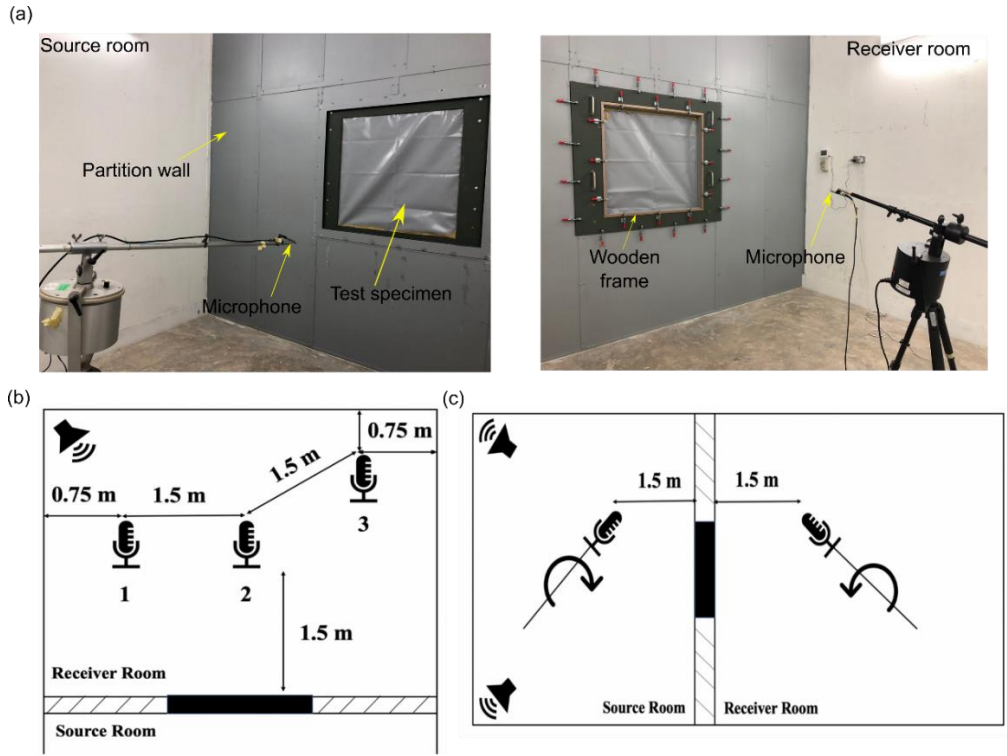

**Figure 3.** (a) The photographs of the installed test specimen for acoustical measurements. Schematic representation of experimental setup for the measurements of (b) decay time T20 and (c) sound pressure levels.

## Results and Discussion

**Sound transmission loss measurement (small-scale)**

A Four-microphone Impedance tube was utilized to measure the sound transmission loss. A 3D-printed ring-shaped sample holder was used for mounting the test samples inside the impedance tube. For transmission loss measurement of the acoustic curtains, the effect of the sample holder was evaluated first. In the process, the transmission loss of the acoustic curtain was measured under two different conditions; with specimen holder and without specimen holder. In the presence of the sample holder, around 1 dB increments in the transmission loss values were observed over 125-1600 Hz. The sample holder had a negligible effect on the measured STL values for the test specimens. Therefore, all the impedance tube measurements were performed using the sample holder assembly. The measured average transmission loss spectra of the various test specimens are shown in Figure 4. For all the materials at the lower frequencies, the transmission loss response was influenced by the sample's static stiffness. In contrast, the transmission loss dip occurred around the corresponding resonant frequency of the materials. After the resonant dip, the higher transmission loss values were governed by the 'Mass-Law.' As per the mass-law, the transmission loss can be expressed as $\Delta TL = 20\log(fm_s) - 48\ dB$, where $f$ is the frequency (Hz), $m_s$ is the surface mass density ($kg/m^2$). As from the expression, the transmission loss value at a specified frequency is related in direct proportion to the surface mass density of the specimen. As shown from transmission loss spectra (Figure 4), the woolen felt (soft) weigh lightest, followed by woolen felt (stiff) with a calculated surface mass density of 0.213 and 0.252 kg/m², respectively. It, therefore, contributed to the lowest average STL values with up to 1dB and 2.5dB retrospectively. The other woven textiles, including hospital curtain, TANGO curtain, and type I Elephant material, showed a similar trend with an average sound transmission loss of around 3-6 dB over 150-1600 Hz. The poor acoustic performance of these woven materials may be attributed to their geometrical configurations. Woven fabrics involve weaving numerous threads perpendicular and alternatingly, while nonwoven fabrics include bonding fibers through chemical, mechanical or thermal treatment. In general, woven materials are more robust, lightweight, and have higher durability but the porous weaves also promote sound penetration across the fabrics.

Furthermore, relatively higher transmission loss was achieved from the nonwoven specimens, including soft textured liners, pure PVC sheets, and PVC-coated PE fabric. The smooth textured liner was composed of thin polyester knit (0.1 kg/m²) covered on both sides with plasticized fire-resistant PVC foam (0.54 kg/m²). In contrast, the PVC-coated polyester was composed of polyester mesh sandwiched between two thermally fused PVC sheets (see Figure 5). With its closed-cell nature of the PVC foam, the lightweight GRIP material provided significant STL improvement of about 2-11 dB for frequencies between 300-1600 Hz over conventional woven curtain fabrics. This significant difference was attributed to the smooth, impervious surface of PVC-based specimens, from which incoming sound waves got reflected, thus resulted in a higher sound transmission loss.

Moreover, the viscoelastic nature of PVC coating provided enhanced damping of the sound vibrations, resulting in reduced sound transmission across the materials. As depicted in Figure 4, the 100% pure PVC provided the highest overall STL with an increase of 2-5 dB in the 150-200 Hz range and 1-2 dB in the 500-1600 Hz range compared to PVC coated polyester. However, the higher mass density and elasticity of pure PVC sheets constituted heaviness, lower flexibility, and pliability for replacement as conventional curtain material.

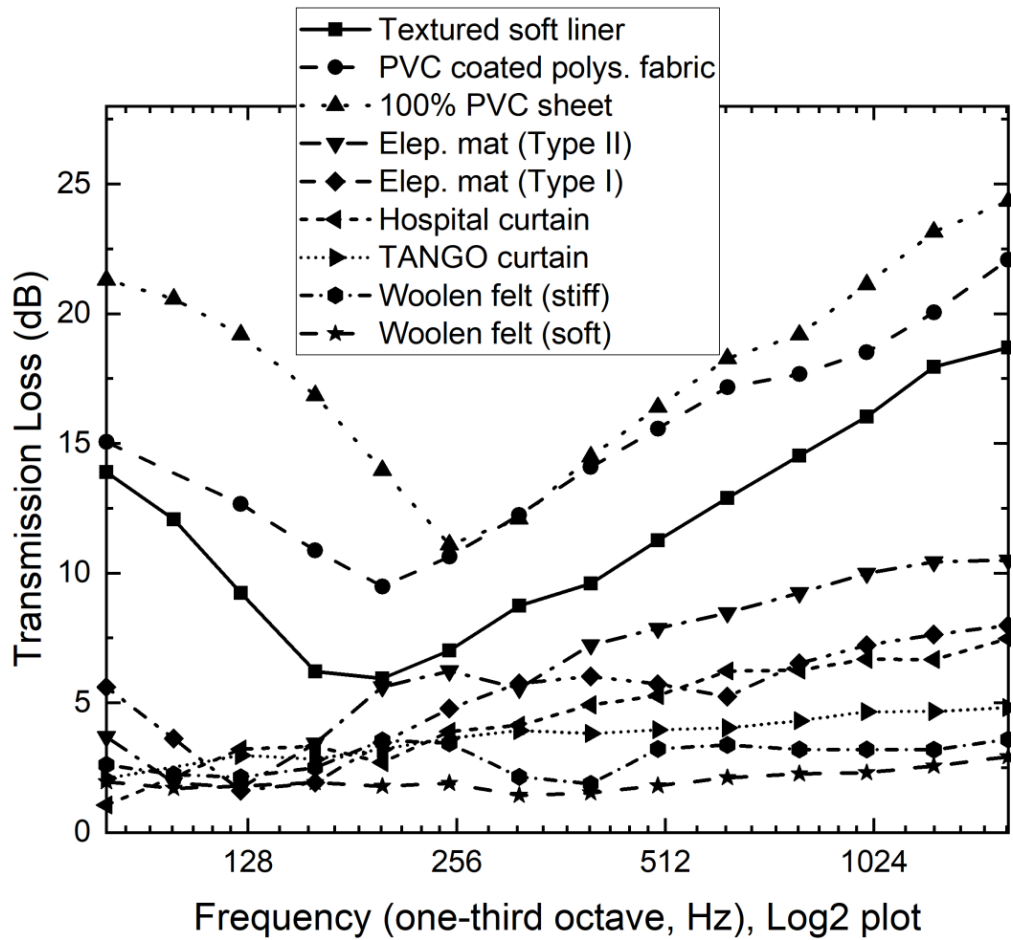

**Figure 4.** Transmission loss performance of various test specimens based on impedance tube test. (a) PVC coated polyester curtain, propylene (PE) sheet, standard hospital curtain, and TANGO acoustical curtain. (b) PVC coated polyester curtain, standard hospital curtain, and 100 % pure PVC. The results were plotted after the exclusion of STL values for the sample holder.

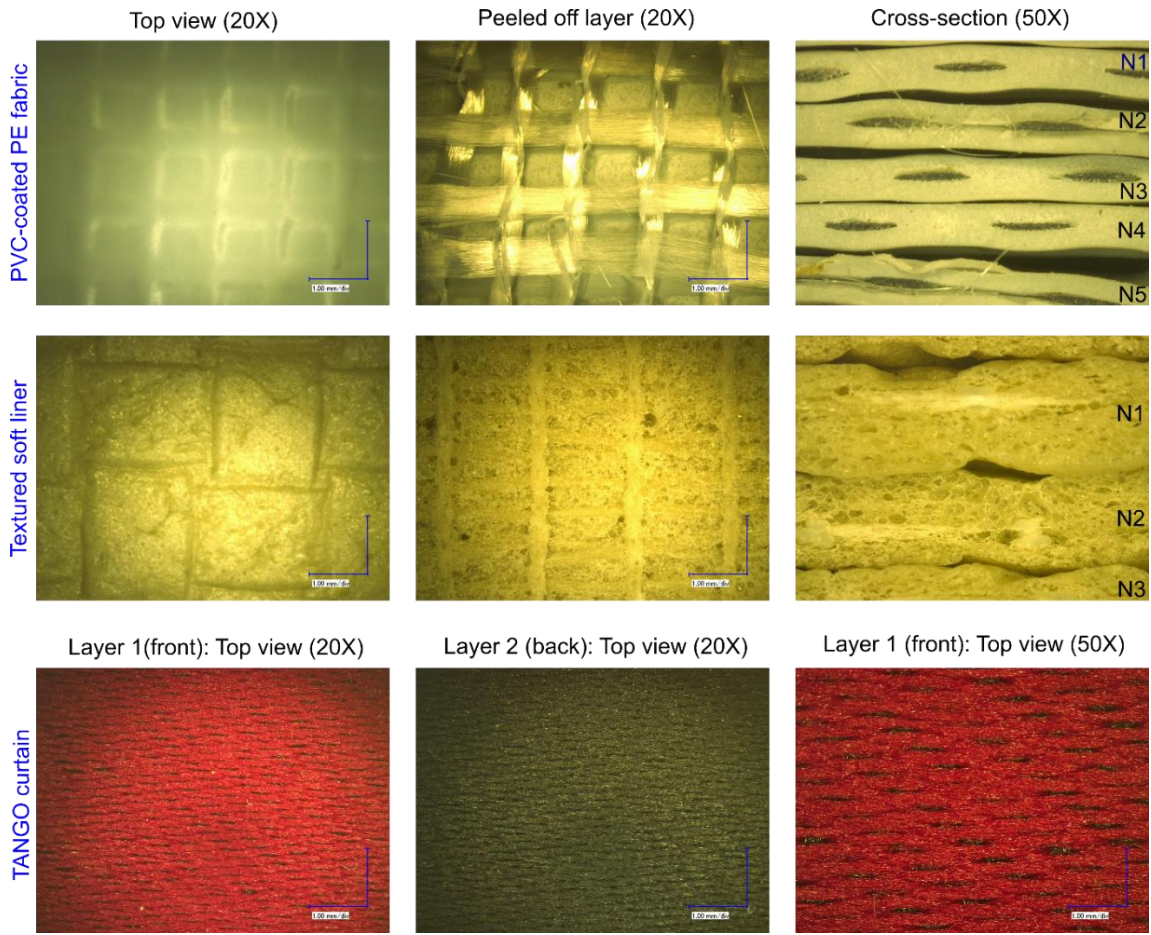

**Figure 5.** Microscopic images of PVC-coated PE fabric, Textured soft liner, and TANGO curtains at different magnifications. Cross-sectional images were captured using the curtains with multiple folding (N1, N2, .., etc.).

**Sound insertion loss measurements (large-scale)**

The impedance tube measurement data (see Figure 4) shows that the soft textured liners (TSL), PVC coated polyester, and pure PVC curtains exhibited higher average transmission loss values in 150-1600 Hz. Therefore, these materials were further investigated in the reverberation room. The curtain with a specified dimension (1m×1m) was installed at the partition wall, and the sound pressure levels were evaluated in a pink noise source setting. The experimental setup is schematically presented in Figure 3. The measured sound insertion loss spectra obtained from these acoustic curtains are shown in Figure 6. The line plots suggested consistency and repeatability with the impedance tube measurements in ranking materials based on their transmission loss performance. The IL values increased with the increment of frequency values following the mass-frequency law. The PVC coated PE fabric's insertion loss (IL) performance and pure PVC sheet curtain were the best-selected materials. The trend was consistent with the results reported by Wang et al. [12] for the sound performance of PVC composite materials. For PVC-based curtains, the average insertion loss values were ranged from 8 to 15 dB in 500-1600 Hz with a peak value of 11 dB for pure PVC and 8.5 dB for GRIP in mid-frequencies of 500-700 Hz. Such high insertion loss values could be related to the viscoelastic properties of the PVC.

Further, the average insertion loss values measured in the reverberation room for both curtains were almost similar (Figure 6). In contrast, the STL values of the 100 % pure PVC sheet were significantly higher than that of the PVC-coated PE curtain (Figure 4). The deviations in the impedance tube trends and the reverberation chamber methods were attributed to the sound source field. In the impedance tube method, an incident one-dimensional plane wave was used as an input source, and the anechoic end was used to minimize the backscattered wave on the downstream side of the tube. A diffuse sound field was generated in the source room for the reverberation room methods, and the sound pressure level was measured in the receiving room. The diffused sound field seems to be a more realistic emulation of the actual ambient condition. Therefore, the reverberation room method provided practical and feasible information about the acoustic performance of test specimens.

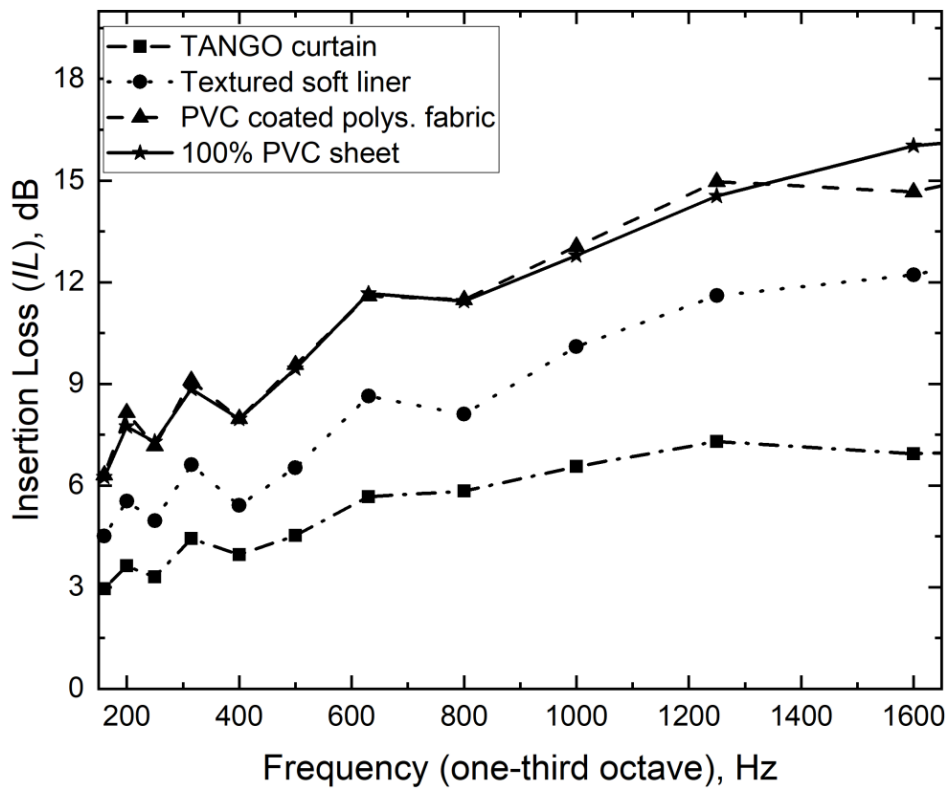

**Figure 6.** Sound insertion loss performance of TANGO curtain, textured soft liner (TSL), PVC-coated polyester fabric, and 100 % pure PVC, measured in the reverberation room.

**Sound insertion loss measurements (Multi-layer curtains)**

While the impedance tube measurements concluded that conventional woven material exhibited relatively poor transmission loss performance, a comprehensive study was conducted to investigate any improvements by inserting a single sheet of PVC-based material between two layers of woven fabrics. As woven materials are generally of lower surface mass density and permeable to sound, perhaps STL could be improved by inserting a sheet of reflective, waterproof, or viscoelastic material in between. As a proof of concept, the tango curtain was utilized as a pocket with a provision to insert the sheet of PVC-based material from the top, as illustrated in Figure 7(a). All measurements were completed similarly to the single material setup with five repetitions in the reverberation room. The IL data was concurrently ensured with the mass law for any improvement.

Figure 7(b) shows the measured sound insertion loss (IL) spectra from inserting various PVC-based materials inside the TANGO curtain. An estimated 1-2 dB and 2-5 dB improvements were observed for PVC coated PE and 100% PVC, respectively, compared to the standalone material. According to mass law, STL should be increased if the inserted material contributes to a more significant overall surface mass density. While the insertion loss of PVC coated PE and pure PVC as standalone materials showed good noteworthy performances, it was observed that pure PVC had a higher overall insertion loss of about 1.5-4.5 dB than PVC coated PE when inserted between the TANGO fabric, as shown in Figure 7(b). One possible explanation for this improvement could be that probable air cavities between the semi-permeable TANGO fabric and impermeable PVC-based material induced internal sound reflection. Since pure PVC has a higher surface mass density than PVC-coated PE, inserting it between the layers of TANGO fabric could have amplified the overall IL performance. Also, there was no noticeable improvement in IL for a soft textured liner (GRIP) through inserting in the TANGO curtain. With a calculated surface mass density of 0.644 kg/m$^2$, GRIP does not contribute significantly to the overall mass density with TANGO (0.224 kg/m$^2$). Another possible explanation could be that its PVC foam coating on soft textured liner may not be as prominent as the PVC-coated PE fabric or pure PVC.

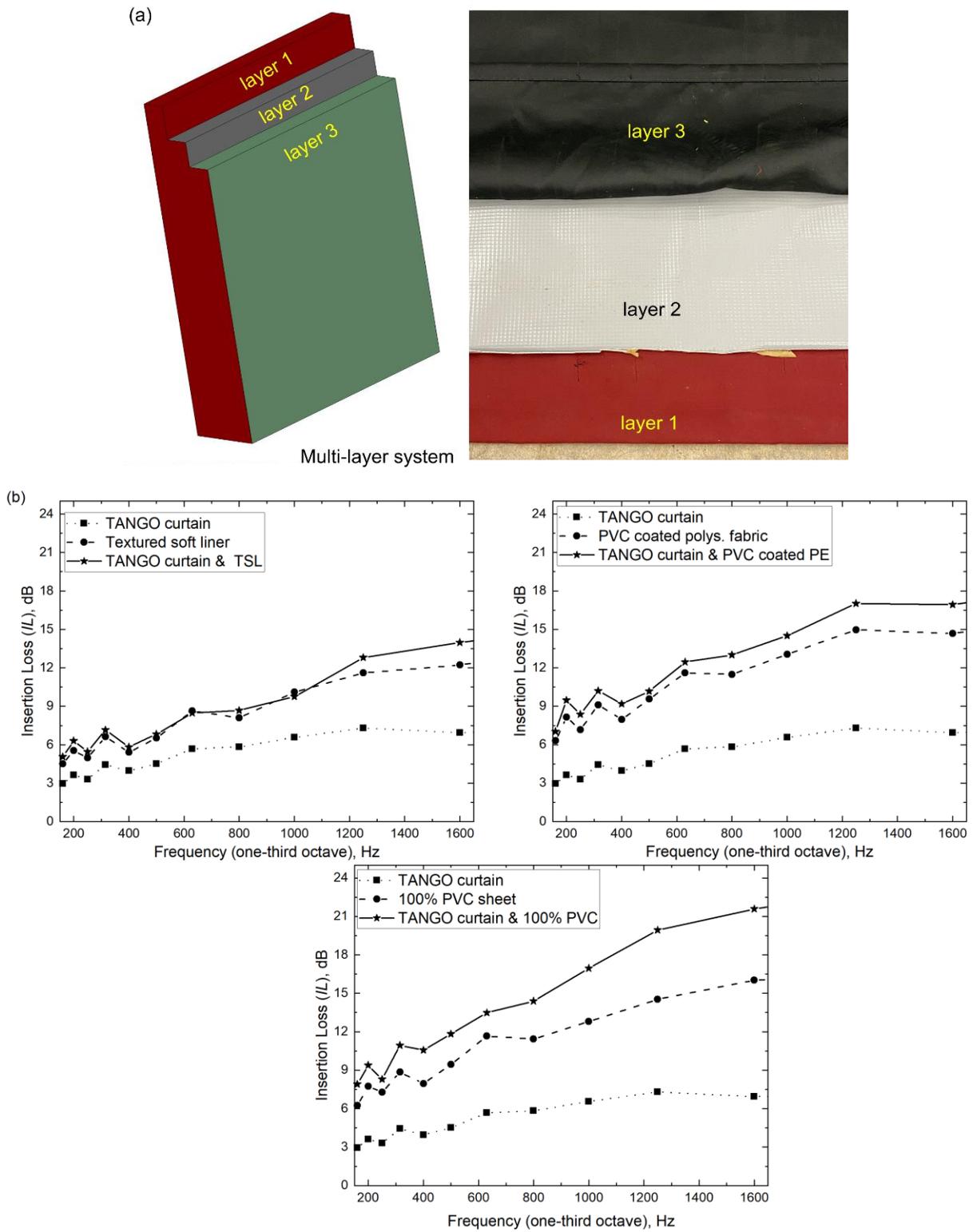

**Figure 7.** (a) Multi-layer curtain schematics and photographs. (b) Measured insertion loss performance of the multi-layer curtain system.

**Conclusion**

In summary, the transmission loss of the curtains was measured using the impedance tube method. Then the reverberation room method was used to measure the insertion loss of the larger samples. As presented earlier, the PVC-based curtains exhibited outstanding sound barrier performance. The average STL values of 100 % pure PVC curtain were measured to be maximum for impedance tube measurement, then PVC-coated PE curtains and PVC-coated textured soft liners. The acoustic transmission loss of acoustic curtains was greatly dependent on the material's surface mass density, wherein the performance was better for the specimen with high surface mass density.

For reverberation room methods, the PVC coated PE fabric's IL performance and pure PVC sheet curtain showed a good agreement. For PVC-based materials, an average insertion loss value of around >10 dB was achieved in the 500-1600 Hz frequency range. Such high insertion loss values attained from the PVC-based curtains are promising and may find a potential application for mid-to-high frequency noise insulation.

Furthermore, the multi-layer curtains improved insertion loss performance of 1-2 dB and 2-5 dB for PVC coated PE and 100% PVC, respectively, compared with the single-layer curtains. The investigations imply that multi-layer curtain designs can be used as acoustic window curtains to reduce outdoor road traffic noise while enabling daylight transmission into the rooms.

**Conflicts of Interest**

The authors would like to declare that they have conducted the acoustic measurements with similar methodologies reported in their previous preprint article (arXiv:2008.06690).